\documentclass[runningheads]{llncs}

\usepackage{llncsdoc}
\usepackage{graphicx}
\usepackage{booktabs}
\usepackage{arydshln}
\usepackage{multirow}
\usepackage{multicol}
\usepackage[FIGTOPCAP,scriptsize]{subfigure}
\usepackage{caption}
\usepackage{amssymb}
\usepackage{bbding}
\begin{document}

\title{FocusNet: Imbalanced Large and Small Organ Segmentation with an End-to-End Deep Neural Network for Head and Neck CT Images}
\titlerunning{FocusNet: Imbalanced Organ Segmentation for HaN CT Images}

\author{
 Yunhe Gao\inst{1}\textsuperscript{*},
 Rui Huang\inst{1}\textsuperscript{*},
 Ming Chen\inst{4,5,6},
 Zhe Wang\inst{1},
 Jincheng Deng\inst{3},
 Yuanyuan Chen\inst{4,5,6},
 Yiwei Yang\inst{4,5,6},
 Jie Zhang \inst{4,5,6},
 Chanjuan Tao \inst{4,5,6},
 \and Hongsheng Li\inst{1,2(}\Envelope\inst{)} }

 %index{Gao, Yunhe}
 %index{Huang, Rui}
 %index{Chen, Ming}
 %index{Wang, Zhe}
 %index{Deng, Jincheng}
 %index{Chen, Yuanyuan}
 %index{Yang, Yiwei}
 %index{Zhang, Jie}
 %index{Tao, Chanjuan}
 %index{Li, Hongsheng}

\authorrunning{Y Gao et al.}

\institute{SenseTime Research, \email{\{claudegyh@gmail.com, huangrui@sensetime.com\}}
\and The Chinese University of Hong Kong, \email{\{hsli@ee.cuhk.edu.hk\}}\\ $^3$
 Shenzhen Yino Intelligence Techonology,\\ $^4$ Inst. of Cancer Research \& Basic Medical Sciences of Chinese Academy of Sciences, \\ $^5$ Cancer Hospital of University of Chinese Academy of Sciences, \\ $^6$ Zhejiang Cancer Hospital }

\maketitle

 \renewcommand{\thefootnote}{\fnsymbol{footnote}}
 \footnotetext[1]{Yunhe Gao and Rui Huang contributed equally to this work.}
\begin{abstract}
In this paper, we propose an end-to-end deep neural network for solving the problem of imbalanced large and small organ segmentation in head and neck (HaN) CT images. To conduct radiotherapy planning for nasopharyngeal cancer, more than 10 organs-at-risk (normal organs) need to be precisely segmented in advance. However, the size ratio between large and small organs in the head could reach hundreds. Directly using such imbalanced organ annotations to train deep neural networks generally leads to inaccurate small-organ label maps. We propose a novel end-to-end deep neural network to solve this challenging problem by automatically locating, ROI-pooling, and segmenting small organs with specifically designed small-organ sub-networks while maintaining the accuracy of large organ segmentation. A strong main network with densely connected atrous spatial pyramid pooling and squeeze-and-excitation modules is used for segmenting large organs, where large organs' label maps are directly output. For small organs, their probabilistic locations instead of label maps are estimated by the main network. High-resolution and multi-scale feature volumes for each small organ are ROI-pooled according to their locations and are fed into small-organ networks for accurate segmenting small organs. Our proposed network is extensively tested on both collected real data and the \emph{MICCAI Head and Neck Auto Segmentation Challenge 2015} dataset, and shows superior performance compared with state-of-the-art segmentation methods.

\end{abstract}
\section{Introduction}

In this paper, we attempt to solve the segmentation problem of imbalanced large and small organs in head and neck (HaN) CT images, where the key challenge is to precisely segment small organs whose volume is much smaller than the average. In HaN radiotherapy planning task, it is of vital importance to accurately determine the locations and volumes of organs-at-risks (OARs). Oncologists would design radiotherapy plans such that the radiation can be concentrated on the lesion area without damaging normal organs. Currently, all OARs are manually annotated by oncologists, which is time consuming, tedious and prone to have high inter- and intra-observer variations. Therefore, a computer-aided head organ segmentation system could significantly lower the work load of doctors.

The main difficulty of the task is the severe imbalance between large and small organs (e.g., the smallest organ, \emph{lens}, only occupy 0.0028\% of the whole 3D volume, while the \emph{parotid gland} is over 250 times larger than \emph{lens}). State-of-the-art segmentation neural networks trained based on samples' natural frequencies would have poor performance on the small organs. In addition, due to the limitation of CT technology and the complex anatomical structure of the human head, the contrast between organs and their surroundings is often low. All these factors coupled together make it difficult to develop a method for segmenting both small and large organs simultaneously and accurately.

Over the past decade, many approaches were proposed to resolve the challenging problem of HaN organ segmentation. Early approaches include atlas-based methods, active contours, graph cut and etc. Atlas-based methods were commonly used where there is only a small number of annotated images available. However, atlas-based methods are based on image registration techniques and might generate incorrect organ maps if the organs are occupied by tumors.

Recently, convolutional neural networks (CNN), with its powerful feature representation capability, have made revolutionary progress in many tasks. 2D CNN models such as U-Net \cite{ronneberger2015u}, and its 3D variants have achieved large performance gain in 2D and 3D segmentation than traditional methods. For OAR segmentation, Ibragimov et al. \cite{ibragimov2017segmentation} proposed the first deep learning-based algorithm. Ren et al. \cite{ren2018interleaved} proposed a interleaved 3D-CNN for segmenting small organs in HaN, where the region of interest is obtained by registration. Zhu et al. \cite{zhu2018anatomynet} proposed a 3D Squeeze-and-Excitation U-Net for fast segmentation. However, existing segmentation CNNs are not optimized for imbalanced organ segmentation, these networks generally produce accurate segmentation maps for large organs, while the accuracy of small organs is often sacrificed.%, or use patch-based methods and registration for region proposal.

To address the issue of imbalanced large and small organ segmentation, we observed how oncologists annotate OARs. For organs with large volume, they usually annotate them at the normal scale, while for the small organs, they first find the location and then zoom in to accurately mark them. According to this observation, we propose a novel end-to-end 3D convolutional neural network, FocusNet, which is delicately designed for accurate segmentation of both large organs and small organs.

The whole network has three parts: main segmentation network (S-Net), Small-Organ Localization branch (SOL-Net), Small Organ Segmentation branch (SOS-Net). S-Net is a strong backbone network, which is responsible for the segmentation of all large organs and also provides features for small-organ segmentation. SOL-Net is trained for highlighting the center locations of the small organs. Based on the results of SOL-Net, high-resolution feature volumes concatenated with multi-scale features volumes are ROI-pooled and fed into to the SOS-Net for fine segmentation of small organs. All the three networks share feature volumes and are jointly optimized. In addition, a weighted focal loss \cite{lin2017focal} combined with generalized dice loss is also used to better deal with the severe sample imbalance problem. The proposed method was evaluated on two datasets: a self-collected HaN dataset with 50 CT scans, and MICCAI 2015 Head and Neck Auto Segmentation Challenge dataset. Our proposed algorithm outperforms state-of-the-art methods on HaN organ segmentation with a large margin.

\section {Method}

\begin{figure}[hbt!]
\centering
  \includegraphics[height=4.5cm]{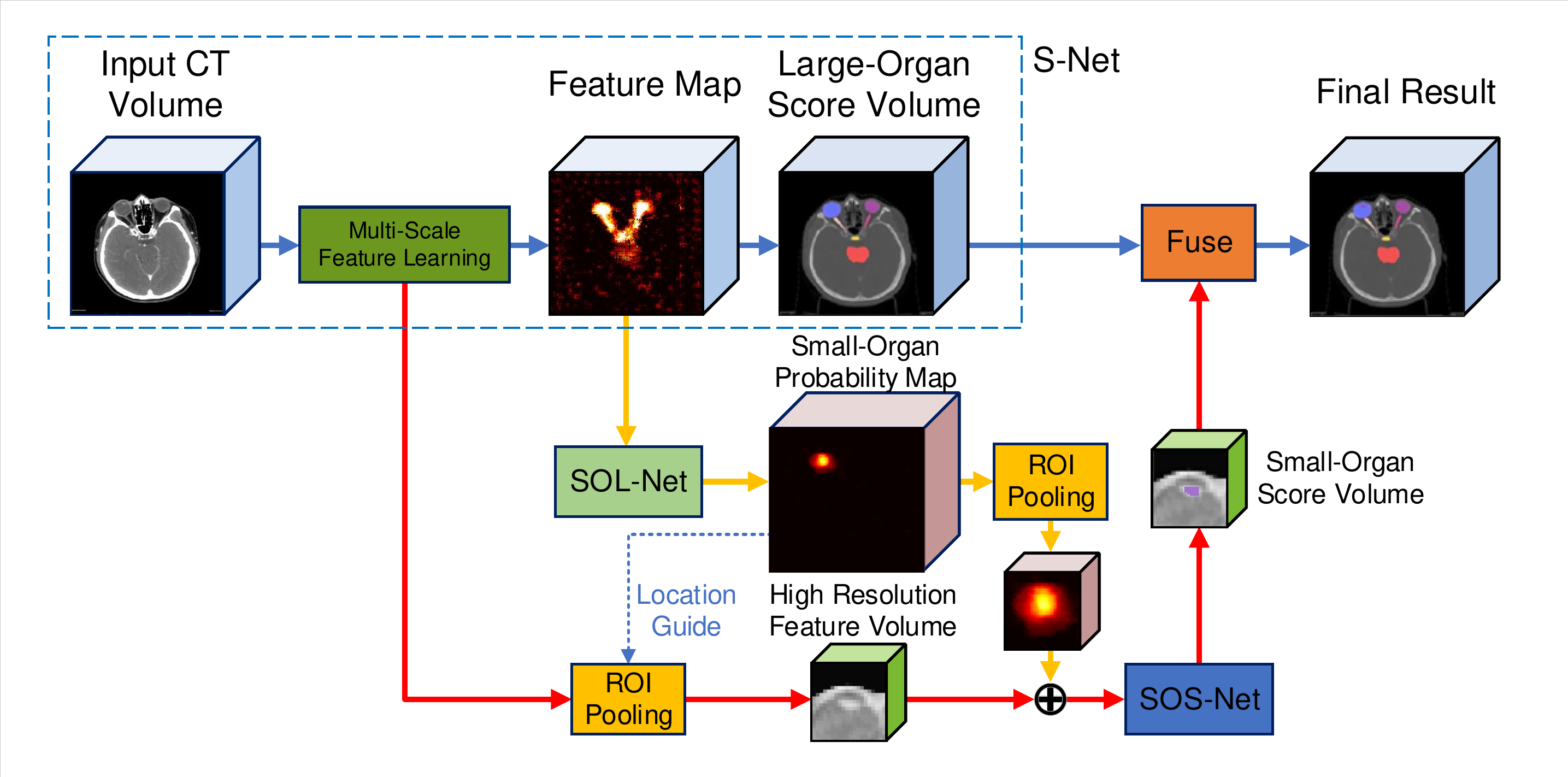}\\
   \setlength{\belowcaptionskip}{-9ex} %caption 和下方文字的距离
   \caption{Overall network structure of our proposed FocusNet.}\label{fig:overview}
\end{figure}

The overall structure of the proposed FocusNet is illustrated in Fig. \ref{fig:overview}. The FocusNet first segments large organs with main segmentation network (S-Net) and localizes the small-organ center locations with Small-Organ Localization branch (SOL-Net). Multi-scale features and high resolution feature are ROI-pooled for small organs to generate small-organ label maps. Therefore, the network decouples localization and segmentation of small organs, and solves the sample imbalance problem, which makes the segmentation of small organs much more easier. The results of large and small organs are then fused to generate the final segmentation label maps.

\subsubsection {Main Segmentation Network (S-Net).}
U-Net is a commonly used 2D CNN, many studies show that its 3D variants have better representation capability in 3D images as they can better capture volumetric contextual information. However, vanilla 3D U-Net has poor performance in OAR segmentation, we address this problem from two aspects. First, U-Net embeds high-resolution information into feature maps by four down-sampling operations, while the decoder reconstructs spatial resolution and obtain dense predictions by deconvolution operations. However, too many times of down-sampling leads to the loss of high-resolution information, which will have catastrophic effects on the small organs that only occupy a few voxels. The shortcut connection between encoder and decoder can only slightly mitigate this problem. Second, UNet can only capture features in fixed scales by downsampling, which limits its representation capability.

\begin{figure}[bt!]
\centering
  \includegraphics[height=3.5cm]{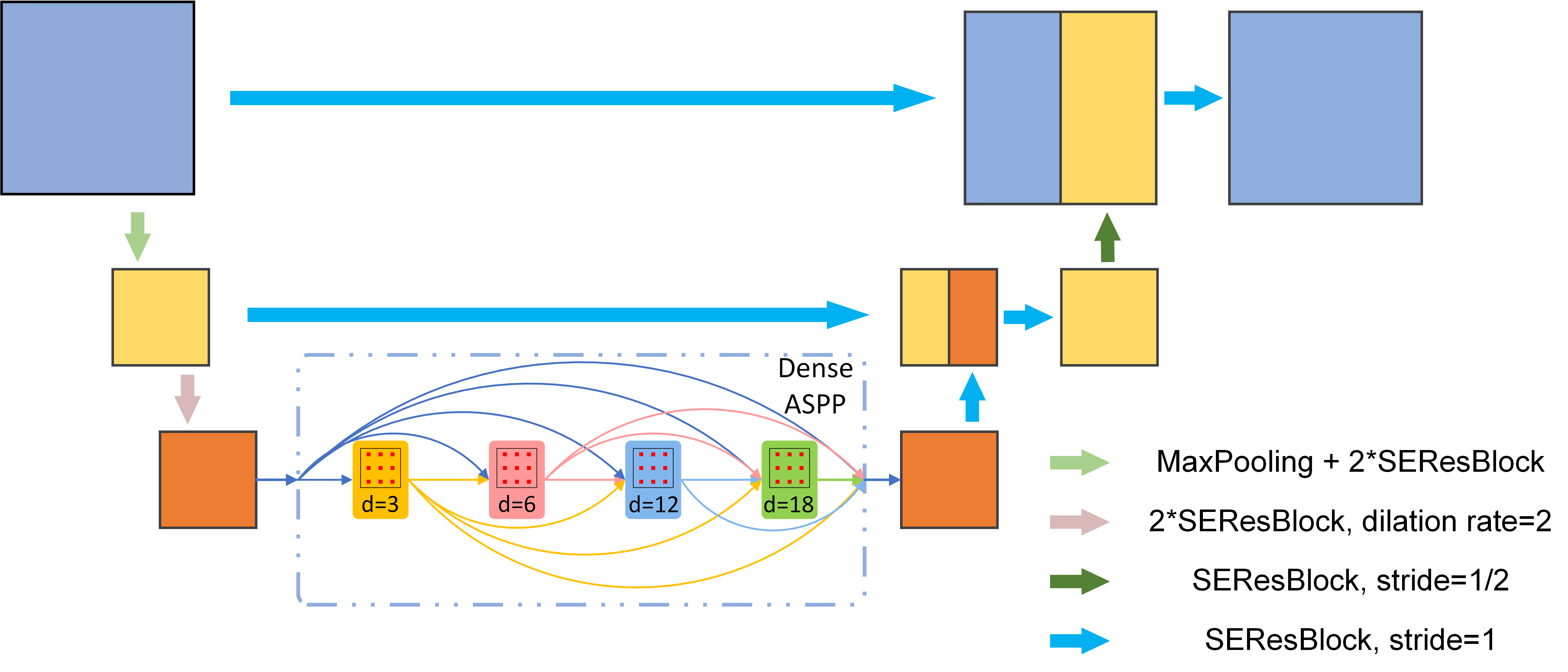}
   \setlength{\belowcaptionskip}{-5ex} %caption 和下方文字的距离
   \captionsetup{font={small}}
   \caption{ Structure of multi-scale feature learning in S-Net. The squares denote feature volumes, and the side length indicates the feature volume size. The \emph{d} in dense ASPP denotes the dilation rate of each convolution kernel.}\label{fig:S-Net}
\end{figure}

The S-Net is elaborately designed to solve the problem mentioned above. As shown in in Fig. \ref{fig:S-Net}, S-Net has a strong backbone, which is a variant of 3D U-Net with residual connections. Squeeze-and-excitation modules \cite{hu2017squeeze} are used for channel-wise attention. To solve the first problem, the S-Net only performs down-sampling once. However, such structure has a disadvantage that the receptive field of convolution kernel is limited, which makes it difficult to integrate global image patterns to learn high-level features. Therefore, dilated convolution and densely connected atrous spatial pyramid pooling (DenseASPP) \cite{yang2018denseaspp} are utilized in our S-Net. DenseASPP can be seen as the combination of the serial connected and parallel connected counterpart, which has the ability of combining arbitrary scales of features through adjusting dilation rate, and better feature reuse.

\subsubsection {Small-Organ Localization Network (SOL-Net).}
 We then mimic the way oncologists annotate the small organs, we propose to design an SOL-Net to first localize the center locations of small organs. As shown in Fig. \ref{fig:overview}, the feature volumes from the last layer of decoder of our S-Net is used as the input of SOL-Net. The training targets are the small-organ center location heat maps, which are created as 3D Gaussian distributions located at the center locations and each small organ has a separate map. The SOL-Net is trained to predict such location maps with a Mean-Square-Error loss. The SOL-Net consists of 2 Squeeze-and-Excitation Residual Blocks (SEResBlock) and a final $1 \times 1 \times 1$ convolution layer with sigmoid layer to output the small-organ location probability maps.

\subsubsection{Small-Organ Segmentation Network (SOS-Net).}
Given the center locations of the small organs from the SOL-Net outputs, we further improve the segmentation accuracy by focusing on the surrounding regions of small organs. Specifically, we first identify the voxel with the highest location probability value from SOL-Net as the small-organ center location, and ROI-pool a 3D feature volume around it. An SOS-Net, which contains 2 SEResBlock and a $1 \times 1 \times 1$ convolution layer, is created for each small organ for outputting the binary segmentation maps. The side-length of the ROI is determined as a fixed value, which is three times of the average diameter of the small organ. In this way, the unbanlanced negative and positive sample problem is solved.

In order to make the best use of all available information, multi-scale feature volumes from the last layer of the S-Net decoder, raw image, and high resolution feature volumes from the first layer of the S-Net encoder are ROI-pooled from the small-organ ROI and concatenated together as the input of SOS-Net. Furthermore, the small-organ location probability heatmap is also concatenated as the spatial location prior. Intuitively, the multi-scale feature volumes from S-Net already encode small organs' segmentation results and the high-resolution feature volumes can help refine the segmentation results. Finally, we integrate the small-organ segmentation results and the large-organ segmentation results to obtain the final prediction for all organs.

\subsubsection{Loss function.}
In our task, the ratio between the background and the smallest organ can reach nearly ${10^5:1}$, which makes the loss dominated by large numbers of easy background samples. We use a focal loss for multi-class classification to solve this problem, and further use weights to balance between organs,

\begin{eqnarray}
L_{Focal}=\sum_{t=0}^{C}-\alpha_t(1-p_t)^\gamma \log(p_t),
\end{eqnarray}
where C is the number of categories, $p_t$ is the probability of class $t$, $\alpha_t$ is the weight of each organ, which is inversely proportional to each organ's average size. ${(1-p_t)^\gamma}$ is the modulating factor which weights less on easy samples (voxels), whose prediction confidence $p_t$ is close to 1. In our experiment, $\gamma$ is set as 2.

Generalized dice loss is another loss function that directly optimize for the evaluation metrics. We take generalized dice loss as the following form:

\begin{eqnarray}
L_{Dice}=\sum_{t=0}^{C}(1 - 2\frac{\sum y_tp_t}{\sum y_t + \sum p_t}),
\end{eqnarray}

where the $y_t$ and $p_t$ are the label and probability of class t.
In our experiment, the combination of focal loss and dice loss results in best segmentation accuracy, the total loss is as follow:

\begin{eqnarray}
L_{total}=L_{Focal}+ L_{Dice},
\end{eqnarray}

\section{Experiments}

 The proposed FocusNet was evaluated on two datasets of HaN CT images. The first dataset, denoted as our dataset, consists of 50 collected samples from hospitals, and 18 organs were delineated manually by doctors. We randomly shuffled our dataset and selected 40 samples for training and 10 samples for testing. For fair comparison with state-of-art methods in OARs segmentation, we further evaluated FocusNet on a public dataset. The \emph{MICCAI Head and Neck Auto Segmentation Challenge 2015} dataset, denoted as MICCAI'15 dataset, consists of 38 samples for training and 10 samples for testing, and has 9 organ annotations. Two evaluation metrics are used in this study: Dice score coefficient (DSC) and 95\% Hausdorff Distance (95HD).

Training is performed in three stages. We first train the S-Net, and then train the SOL-Net while fixing the trained parameters of S-Net. The SOS-Net is trained at last because it needs the resulting feature volumes and the location probability maps from the SOL-Net. At last, we end-to-end finetune the whole network for joint optimization.

\subsection{Experiments on our collected dataset}
The average number of voxels of each organ is shown in supplementary material. The sample unbalance problem is severe, where large organs occupies over tens of thousands of voxels while small organs only occupies hundreds or even tens of voxels. Organs with voxels fewer than 1000 are considered as small organs, which results in 10 small organs in total.
\begin{figure}[!tb]
\tiny
\centering
  \subfigure[Atlas]{
    \includegraphics[width=0.17\linewidth]{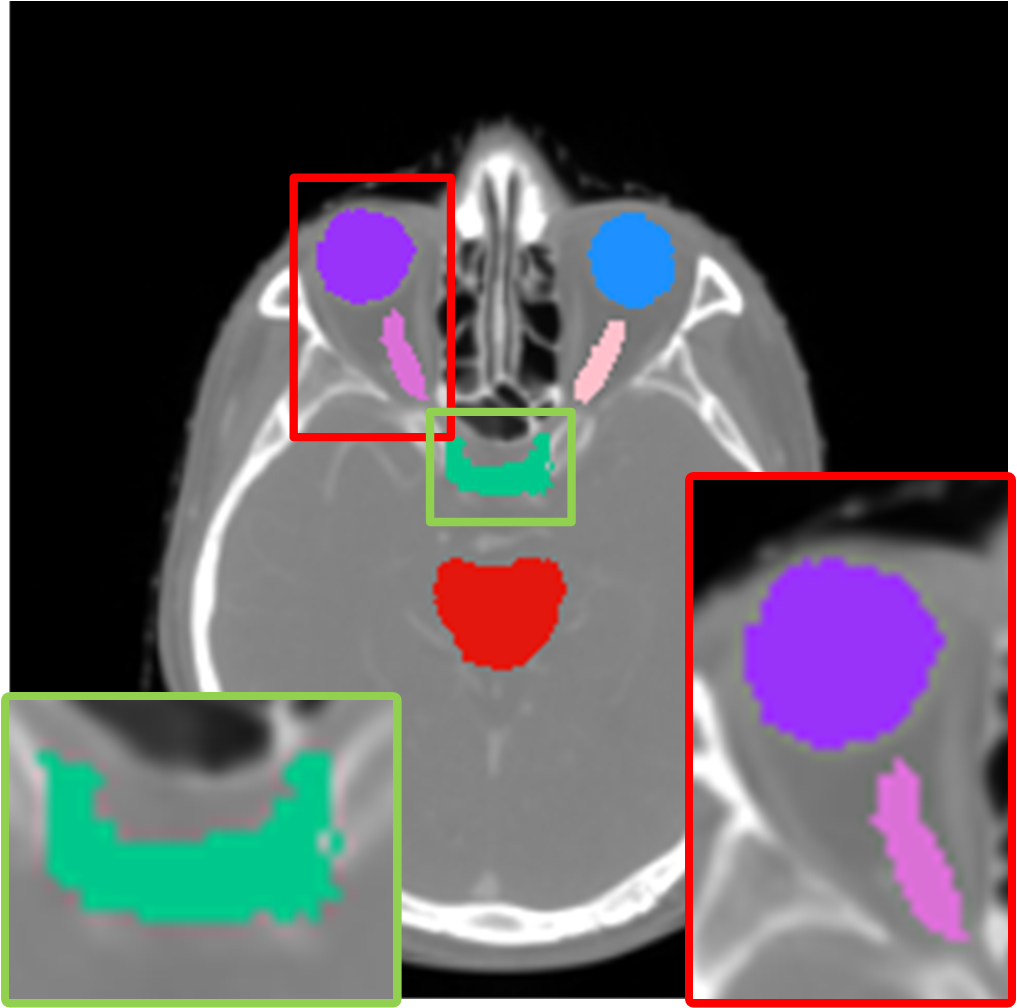}
  }
  \subfigure[SEResUnet]{
    \includegraphics[width=0.17\linewidth]{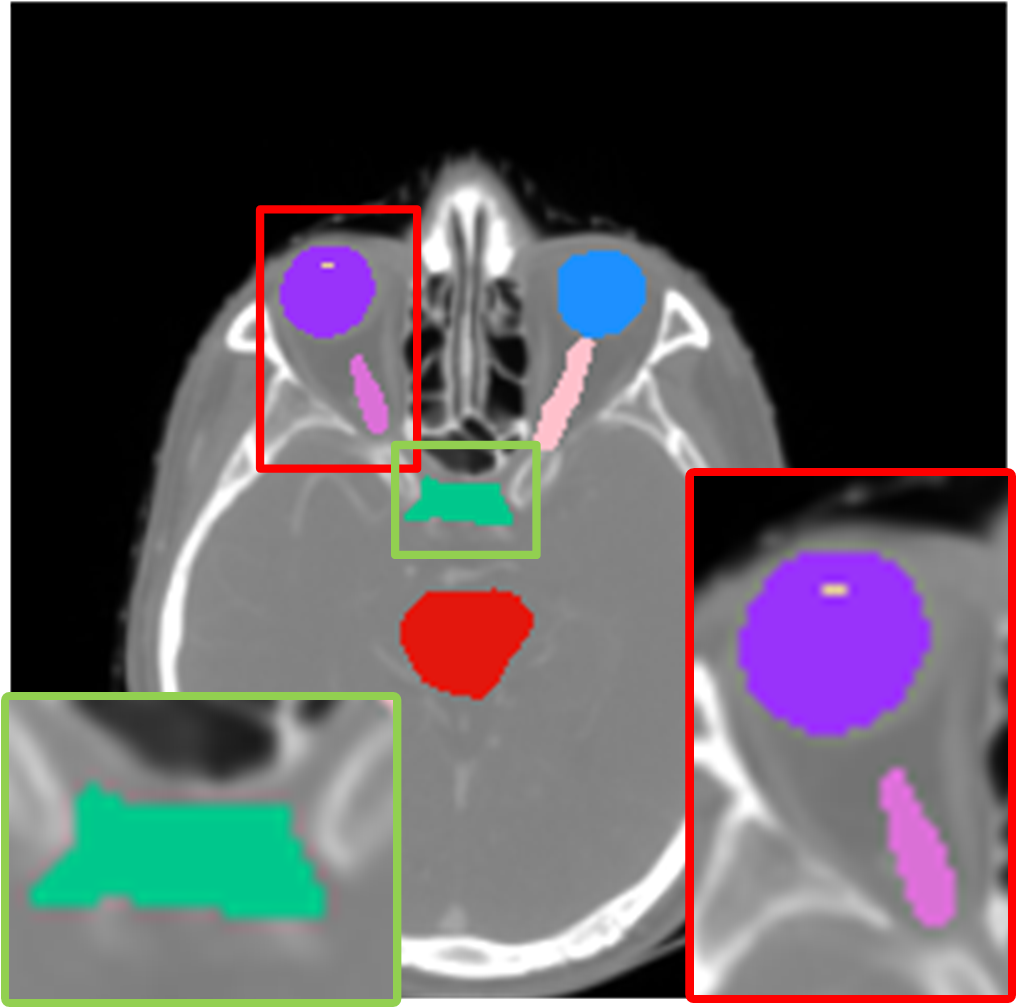}
  }
  \subfigure[DeepLab v3+]{
    \includegraphics[width=0.17\linewidth]{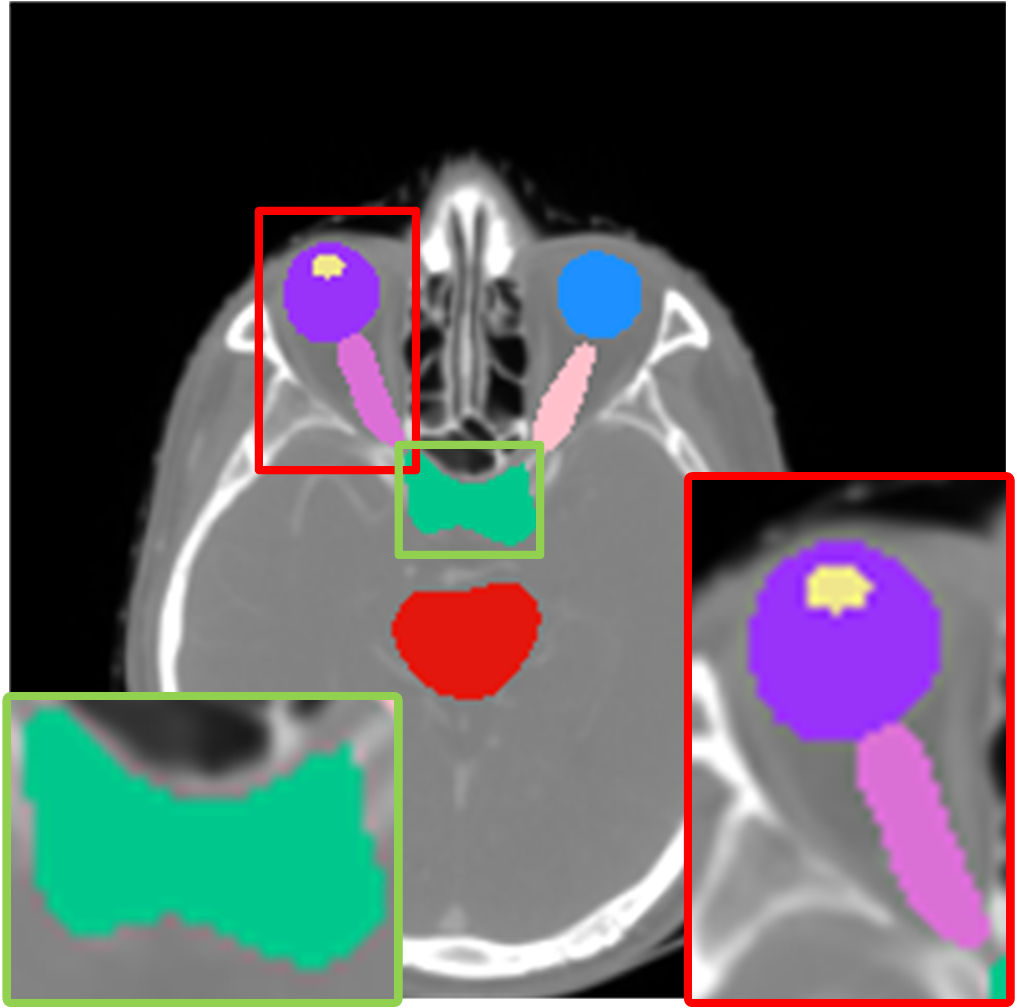}
  }
  \subfigure[FocusNet]{
    \includegraphics[width=0.17\linewidth]{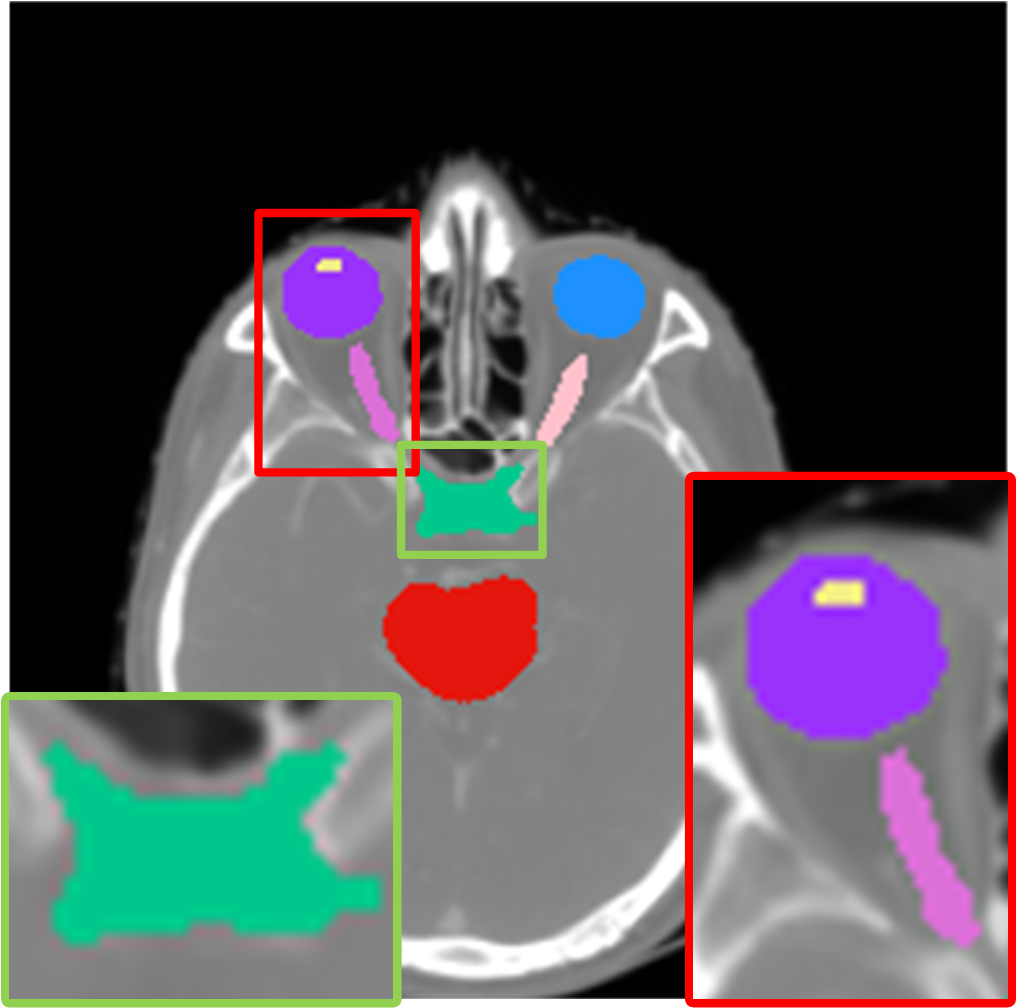}
  }
  \subfigure[GT]{
    \includegraphics[width=0.17\linewidth]{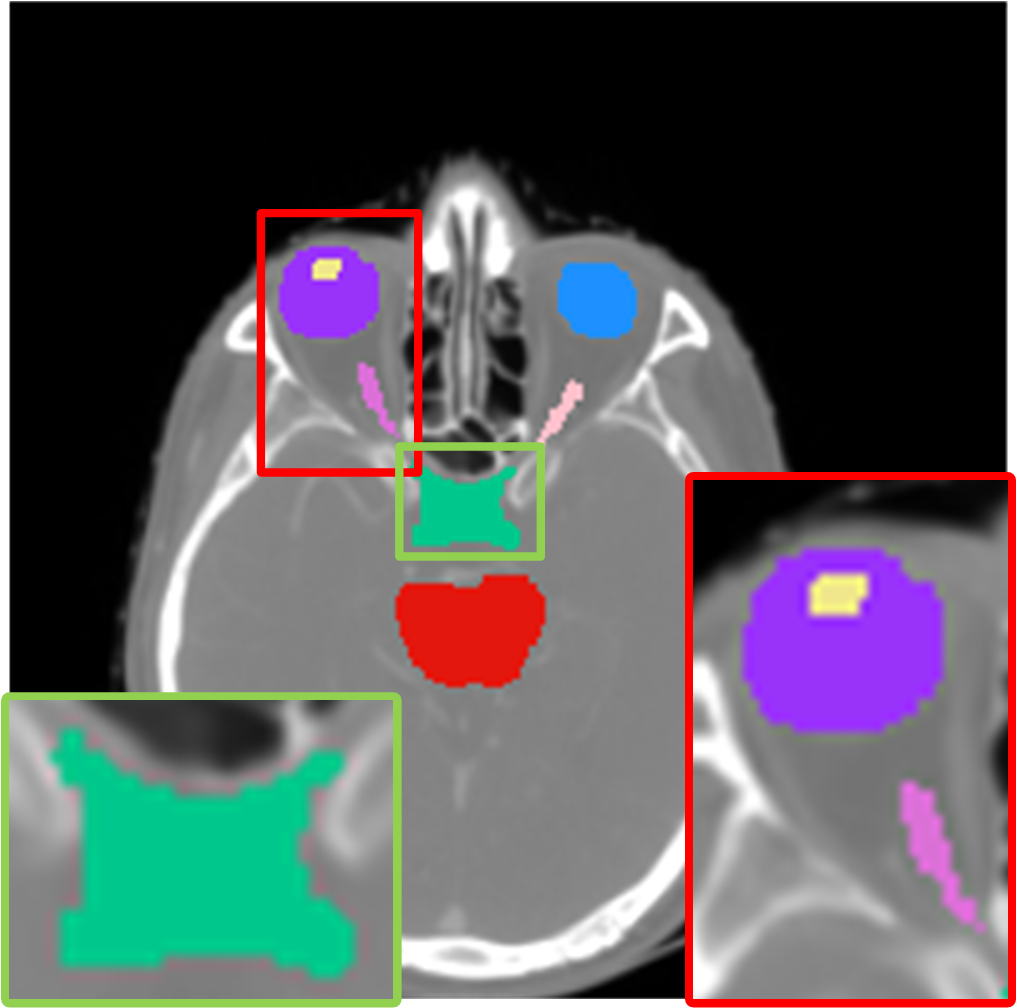}
  }\\
    \setlength{\belowcaptionskip}{-10ex} %caption 和下方文字的距离
    \captionsetup{font={small}}
  \caption{Some segmentation examples of comparative experiment, GT is ground truth.}
  \label{fig:Comparation}
\end{figure}

\subsubsection {Comparison with other methods.}
We compare our proposed method with 3D U-Net \cite{ronneberger2015u}, a 3D variant of DeepLab-v3+ \cite{chen2018encoder}, and an atlas-based method \cite{avants2008symmetric}.
U-Net \cite{ronneberger2015u} is popular in 2D medical image segmentation. We modify the U-Net (denoted as SERes U-Net) to replace original convolution with 3D Squeeze-and-Excitation Residual Blocks (SEResBlock) to increase its segmentation accuracy. DeepLab-v3+ \cite{chen2018encoder} is successful for 2D semantic segmentation, we extend their network structure to 3D for volumetric segmentation. All the compared deep learning based methods were randomly initialized and trained using the same loss function as our proposed FocusNet.
For the atlas-based method, Symmetric Normalization (SyN) \cite{avants2008symmetric}, its implementation in ANTs software package is used to generate template image (atlas) and the template label from training set. When given a CT to be segmented, the optimal transformation between the atlas and target was obtained by registration, and then this transformation is applied to the template label.

Comparative results are shown in Table \ref{tbl:result}. For large organs, deep learning methods have slightly better results, but for small organs, deep learning methods have large advantages than atlas-based method SyN. It is because that small organs have more complex anatomical structures, atlas-based method has limited capability of dealing with complicated and diverse anatomy variations. Among deep learning based methods, our proposed FocusNet outperforms other methods by large margins. This is because it has specific mechanisms for handling small organs and could therefore generate more accurate resulting label maps.

For qualitative comparison, as seen in Fig. \ref{fig:Comparation}, for the optic nerves, the prediction given by DeepLab-v3+ is much larger than the ground truth, FocusNet gives a precise segmentation result. For the optic chiasm, the result of other methods are in a mess, it is difficult to see a clear shape, while FocusNet gives the best segmentation with X-like shape. SERes U-Net has poor results in lens, it is because lens only occupy tens of voxel, too much information is lost due to down-sampling.

\begin{table}[!t]
\tiny
\minipage[t]{0.62\textwidth}
\vspace{0pt}
\centering
\setlength{\tabcolsep}{1.5mm}{
\begin{tabular}[t!]{lcccc}
\toprule
Organs         & Atlas-based            & SERes U-Net          & DeepLab                    & Ours\\\midrule
Brain Stem     & 80.6$\pm$1.7           & 79.2$\pm$1.8           & 84.0$\pm$1.2             & \textbf{85.8$\pm$1.4}    \\
Eye L          & 78.7$\pm$3.5           & 86.1$\pm$2.6           & 86.7$\pm$2.2             & \textbf{87.6$\pm$0.8}     \\
Eye R          & 83.9$\pm$3.4           & 88.2$\pm$1.6           & 90.5$\pm$0.6             & \textbf{91.2$\pm$0.9}    \\
Oral Cavity    & 75.2$\pm$13.1          & \textbf{81.0$\pm$10.6} & 77.7$\pm$16.2            & 79.2$\pm$3.5              \\
Parotid L      & \textbf{80.6$\pm$4.3}  & 75.1$\pm$6.4           & 78.8$\pm$4.0             & 77.2$\pm$4.2     \\
Parotid R      & 80.7$\pm$3.2           & 79.8$\pm$1.7           & \textbf{81.7$\pm$3.0}    & 80.0$\pm$3.1     \\
Larynx         & 64.0$\pm$29.1          & 62.8$\pm$28.9          & \textbf{67.7$\pm$29.6}   & 65.8$\pm$29.3              \\
Spinal Cord    & 82.4$\pm$2.1           & 85.9$\pm$2.0           & 85.2$\pm$2.2             & \textbf{87.4$\pm$1.9}     \\\cdashline{1-5}[2pt/2pt]
Lens L         & 24.0$\pm$8.3           & 60.8$\pm$4.2           & 59.2$\pm$8.4             & \textbf {80.8$\pm$4.7}    \\
Lens R         & 26.9$\pm$1.5           & 57.1$\pm$6.2           & 64.3$\pm$6.6             & \textbf {79.0$\pm$6.4}    \\
Opt. Ner. L    & 47.7$\pm$10.6          & 56.0$\pm$3.7           & 52.7$\pm$9.9             & \textbf {63.9$\pm$3.9}    \\
Opt. Ner. R    & 48.5$\pm$6.0           & 49.3$\pm$9.5           & 57.1$\pm$21.4            & \textbf{61.7$\pm$12.1}     \\
Opt. Chiasm    & 54.8$\pm$9.0           & 54.0$\pm$7.6           & 55.6$\pm$11.3            & \textbf{63.8$\pm$11.4}              \\
Pituitary      & 44.6$\pm$12.0          & 67.6$\pm$12.7          & \textbf{78.1$\pm$10.8}   & 76.9$\pm$7.2              \\
Mid. Ear L      & 56.4$\pm$9.7          & 55.2$\pm$15.6          & 51.9$\pm$25.3            & \textbf{56.7$\pm$16.7}     \\
Mid. Ear R      & \textbf{56.2$\pm$14.5}& 47.4$\pm$13.4          & 46.6$\pm$21.8            & 52.2$\pm$20.9              \\
T.M.J. L   & 46.9$\pm$14.1          & 56.5$\pm$8.1           & \textbf{64.7$\pm$3.9}    & 58.4$\pm$7.3              \\
T.M.J. R    & 50.3$\pm$18.8         & 55.1$\pm$12.1          & \textbf{66.1$\pm$8.4}    & 57.2$\pm$5.6              \\\midrule
%AVG           & 62.3$\pm$9.2           & 70.9$\pm$9.8           & 70.2$\pm$8.8             & \textbf {72.0$\pm$}    \\
Average        & 62.3                   & 66.5                   & 69.4                     & \textbf{72.5} \\\bottomrule
\end{tabular}}
\setlength{\belowcaptionskip}{-10ex} %caption和下方文字的距离
\captionsetup{font={scriptsize}}
\caption{Result of different approaches on the 18 HaN organs, measured by DSC. Below the dash line are small organs.}\label{tbl:result}
\endminipage\hspace{20pt}
\minipage[t]{0.3\textwidth}
\vspace{0pt}
\centering
\setlength{\tabcolsep}{2mm}{
\begin{tabular}[t!]{lc}
\toprule
Method     &AVG \\
\midrule
(All Organs) \\
SERes U-Net CE Loss & 60.2  \\
SERes U-Net 1-ds & 63.5\\
SERes U-Net 2-ds & 63.3 \\
SERes U-Net 3-ds & 61.9\\
SERes U-Net+FL & 64.4\\
SERes U-Net+FL+DL             & 69.2  \\
S-Net & 70.6  \\
FocusNet   & \textbf{72.5}  \\
Fat U-Net &70.5  \\
\hline
(Small Organs Only) \\
S-Net                  & 61.5 \\
FocusNet ROI Size 2x                & 63.9\\
FocusNet ROI Size 3x                  & \textbf{65.0}\\
FocusNet ROI Size 5x                 & 64.5\\
\bottomrule
\end{tabular}}
\setlength{\belowcaptionskip}{-10ex} %caption 和下方文字的距离
\captionsetup{font={scriptsize}}
\caption{Ablation study of FocusNet. Above the midline are experiments about network structure and loss function, below are experiments about ROI size of SOS-Net.}\label{tbl:Ablation}
\endminipage
\end{table}

\subsubsection{Ablation Study of FocusNet.}
\vspace{-1ex}
 The ablation results are shown in Table \ref{tbl:Ablation}. Our baseline is 3D SERes U-Net, which utilizes the SEResBlock but has 4 downsampling operations, and adopts the cross-entropy loss. The baseline results in decent segmentation results for large organs while poor for small organs. We then test the baseline, SERes U-Net, with 1, 2 and 3 down-sampling operation respectively, and the one with 1 down-sampling results in highest performance. Utilization of the focal loss (FL) and then combine the dice loss (DL) improve the segmentation accuracy. Combining the DenseASPP module into the network, which is our S-Net, slightly increases the performance. Our proposed FocusNet boosts the final performance by adopting specifically designed small-organ networks.

 We increase the number of channels of all layers of S-Net by a ratio, named Fat U-Net, so that it has the same parameters with FocusNet, result shows that without solving the organ imbalance problem, more parameters cannot boost the performance. We also conducted experiments to find the best ROI size, which is obtained when the side length of the ROI is 3 times than that of the organ.

\subsection{Experiments on MICCAI'15 dataset}

\begin{table}[!thb]
\tiny
\minipage[t]{0.65\textwidth}
\vspace{0pt}
\centering
\setlength{\tabcolsep}{0.5mm}{
%\resizebox{0.8\textwidth}{30mm}{
\begin{tabular}{lcccccc}
\toprule
\multirow{2}[1]{*}{Organs} &  MICCAI                                   & Ren                     & Wang       & Zhu      & \multirow{2}[1]{*}{S-Net} & \multirow{2}[1]{*}{FocusNet}\\
                           &  2015 \cite{raudaschl2017evaluation}       & et al. \cite{ren2018interleaved}   & et al. \cite{wang2018hierarchical}     & et al. \cite{zhu2018anatomynet}    &                           & \\\midrule
Extra Data            & $\times$               & $\times$       & $\times$                 & $\checkmark$                & $\times$            &$\times$  \\
\cline{1-7}
Brain Stem      & 88.0    & N/A                    & \textbf{90.3$\pm$4} & 86.7$\pm$2                & 86.8$\pm$2.9         & 87.5$\pm$2.6\\
Chiasm          & 55.7    & 58$\pm$17              & N/A                & 53.2$\pm$15                & 57.4$\pm$25.1         & \textbf{59.6$\pm$18.1}\\
Mandible        & 93.0    & N/A                    & \textbf{94.4$\pm$1} & 92.5$\pm$2                & 92.5$\pm$1.5         & 93.5$\pm$1.9\\
Opt. Ner. L    & 64.4     & 72$\pm$8               & N/A                & 72.1$\pm$6                & 71.8$\pm$6.9         & \textbf{73.5$\pm$9.6}\\
Opt. Ner. R    & 63.9     & 70$\pm$9               & N/A                & 70.6$\pm$10                & 71.9$\pm$9.9         & \textbf{74.4$\pm$7.2}\\
Parotid L       & 82.7    & N/A                    & 82.3$\pm$6          & \textbf{88.1$\pm$2}       & 86.1$\pm$2.6         & 86.3$\pm$3.6\\
Parotid R       & 81.4    & N/A                    & 82.9$\pm$6          & 87.4$\pm$4                & 87.8$\pm$4.6         & \textbf{87.9$\pm$3.1}\\
Subman. L       & 72.3    & N/A                    & N/A                & \textbf{81.4$\pm$4}       & 79.4$\pm$9.8         & 79.8$\pm$8.1\\
Subman. R       & 72.3    & N/A                    & N/A                & \textbf{81.3$\pm$4}       & 79.7$\pm$4.5         & 80.1$\pm$6.1\\
Average         & 74.9    & N/A                    & N/A                & 79.25                & 79.24         & \textbf{80.29}\\

\bottomrule
\end{tabular}}
\setlength{\belowcaptionskip}{-10ex} %caption和下方文字的距离
\captionsetup{font={scriptsize}}
\caption{Dice overlap coefficient of independent test set in the MICCAI 2015 dataset.}\label{MICCAI_Dice}
\endminipage\hspace{10pt}
\minipage[t]{0.3\textwidth}
\vspace{0pt}
\centering
\setlength{\tabcolsep}{3mm}{
\begin{tabular}[t!]{lc}
\toprule
\multirow{2}[1]{*}{Method}                      & AVG \\
                                                & 95HD\\
\midrule
MICCAI 2015\cite{raudaschl2017evaluation}       & 4.14 \\
Zhu et al.\cite{zhu2018anatomynet}              & 6.72 \\
FocusNet                                        & 2.62 \\
\bottomrule
\end{tabular}}
\setlength{\belowcaptionskip}{-10ex} %caption 和下方文字的距离
\captionsetup{font={scriptsize}}
\caption{Average 95\% Hausdorff distance of test set of MICCAI 2015 dataset. The complete table can be seen in supplementary material. }\label{MICCAI_HD}
\endminipage
\end{table}

We also test our FocusNet on MICCAI 2015 Head and Neck dataset. All the settings of the FocusNet are the same as those used in experiments on our collected dataset, except the number of small organ branch is set as 3 since only 3 organs meet our previous definition of small organs: left and right optic nerve and optic chiasm. Visualization can be seen in supplementary material.

The evaluation results are shown in Table \ref{MICCAI_Dice} and Table \ref{MICCAI_HD}. We compared the highest score from the top four teams in MICCAI 2015 challenge \cite{raudaschl2017evaluation}. For the result of Zhu et al. \cite{zhu2018anatomynet}, it should be noted that they used 38 samples provided by the MICCAI 2015 Challenge combined with additional 216 samples for training.

Our backbone S-Net achieves state-of-the-art performance. It reaches comparable performance in Dice score to Zhu et al. \cite{zhu2018anatomynet} but with only $15\%$ of training data, which shows that S-Net has stronger feature representation capability. Moreover, S-Net has much better result in terms of 95HD, because outliers are alleviated by enlarging the receptive field. After adding SOL-Net and SOS-Net, FocusNet achieves further improvement, especially for the three small organs.

Wang et al. \cite{wang2018hierarchical} proposed a vertex regression-based method, which has good performance in brain stem and mandible, however, has relatively poorer performance in parotid glands, and they did not provide results of other organs. Compared with the registration-based region proposal and patch-based segmentation method used by Ren et al. \cite{ren2018interleaved}, our approach integrates the localization and segmentation of small organs into a unified deep learning framework, which is much faster, has no redundant computation, and results in better performance.

%However, the registration-based region proposal is time-consuming while the patch-based CNN has redundant calculations.

\section {Conclusion}

We proposed an end-to-end deep neural network, FocusNet, which outperforms state-of-the-art methods on segmentation of imbalanced OARs in HaN CT images. By reducing the number of down-sampling and utilizing multi-scale features learned by DenseASPP, our S-Net can guarantee the accuracy of the segmentation of large organs. Trained for predicting small-organ center location maps, SOL-Net can generate accurate small-organ central locations. SOS-Net can solve the sample imbalance problem, and high-resolution feature volumes can be utilized to accurately segment small organs and thus can further boost the performance. A weighted focal loss combined with dice loss is introduced to mitigate the sample imbalance problem. Extensive experiments on real patients' data and the MICCAI 2015 dataset show the effectiveness of our proposed FocusNet.

\subsubsection{Acknowledgements.} This work has been supported in part by the General Research Fund through the Research Grants Council of Hong Kong under Grants CUHK14208417 and CUHK14239816, in part by the Hong Kong Innovation and Technology Support Programme (No. ITS/312/18FX), in part by National Key R\&D Program of China (2017YFC0113201) and Zhejiang Key R\&D Program (2019C03003).

\bibliographystyle{splncs04}

\newpage

\begin{center}
    \Large{\textbf{Supplementary}}
\end{center}
In the supplementary material, we report a) the statistical results of the number of voxels of each organ on our collected dataset, b) some qualitative results of 95HD compared with state-of-the-art methods in OAR segmentation in MICCAI 2015 dataset, c) some visualization of segmentation results of MICCAI 2015 dataset.

\subsubsection{Experiments on our collected dataset}\
 \begin{figure}[!tbh]\centering
\includegraphics[height=4.5cm]{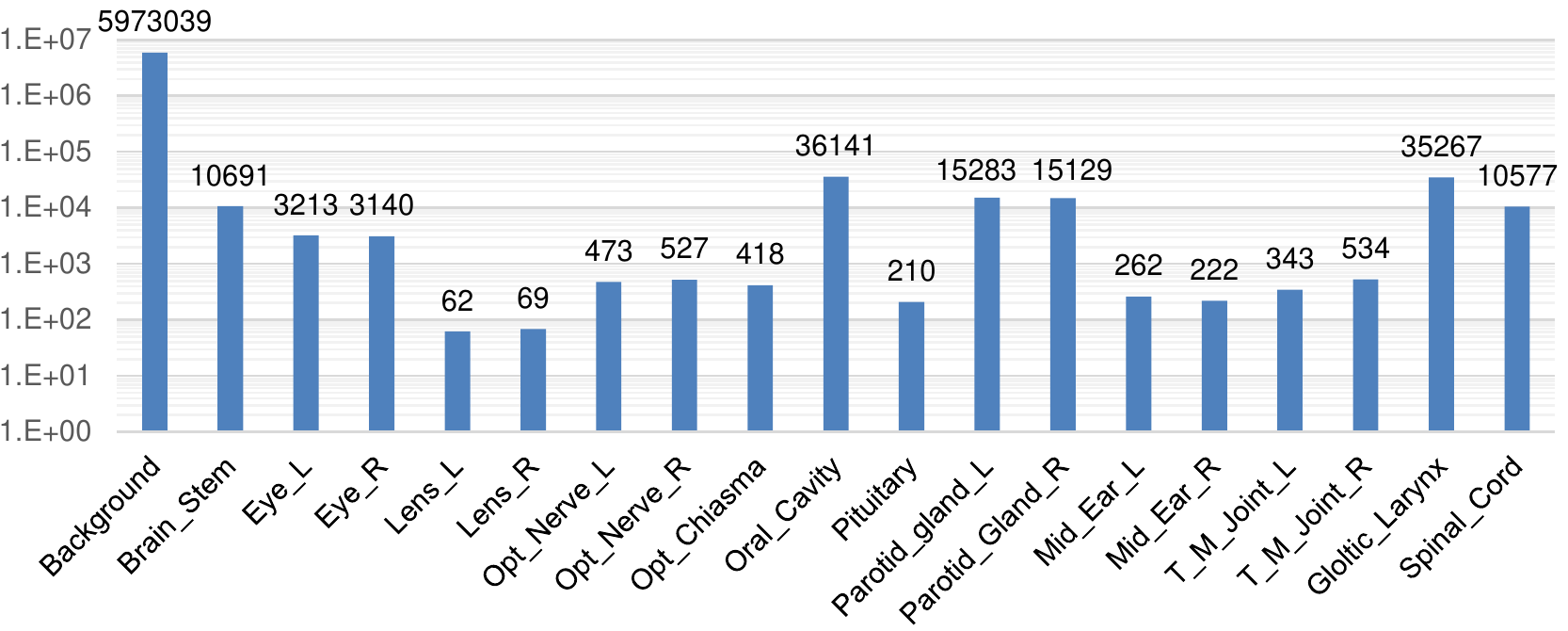}
    \setlength{\abovecaptionskip}{0pt}
  \setlength{\belowcaptionskip}{-7ex} %caption 和下方文字的距离
  \captionsetup{font={normal}}
  \caption{Voxel number of 18 OARs in our collected dataset. The vertical axis is log scale. The sample imbalance problem is severe, where large organs occupy over tens of thousands of voxels while small organs only occupy tens of voxels.
  }\label{fig:class_stastic}
\end{figure}

\subsubsection{Experiments on MICCAI'15 dataset}\

\begin{table}[tbh]

\centering
\setlength{\tabcolsep}{5mm}{
\begin{tabular}[t!]{lcccc}

\toprule
\multirow{2}[1]{*}{Organs} &  MICCAI                               & Ren                                   & Zhu                               & \multirow{2}[1]{*}{FocusNet}\\
                           &  2015\cite{raudaschl2017evaluation}   & et al. \cite{ren2018interleaved}      & et al. \cite{zhu2018anatomynet}      & \\
\midrule
Brain Stem      & 4.59          & N/A                   & 6.42$\pm$2.4        & \textbf{2.14$\pm$0.6}\\
Chiasm          & \textbf{2.78} & 2.81$\pm$1.6          & 5.76$\pm$2.5       & 3.16$\pm$1.3\\
Mandible        & 1.97          & N/A                   & 6.28$\pm$2.2        & \textbf{1.18$\pm$0.3}\\
Opt. Ner. L     & 2.76          & \textbf{2.33$\pm$0.8} & 4.85$\pm$2.3        & 3.76$\pm$2.9\\
Opt. Ner. R     & 3.15          & \textbf{2.13$\pm$1.0} & 4.77$\pm$4.3        & 2.65$\pm$1.5\\
Parotid L       & 5.11          & N/A                   & 9.31$\pm$3.3        & \textbf{2.52$\pm$1.0}\\
Parotid R       & 6.13          & N/A                   & 10.08$\pm$5.1       & \textbf{2.07$\pm$0.8}\\
Subman. L       & 5.35          & N/A                   & 7.01$\pm$4.4        & \textbf{2.67$\pm$1.3}\\
Subman. R       & 5.42          & N/A                   & 6.02$\pm$1.8        & \textbf{3.41$\pm$1.4}\\
Average         & 4.14          & N/A                   & 6.72                & \textbf{2.62}\\
\bottomrule
\end{tabular}}
\setlength{\belowcaptionskip}{-9ex} %caption 和下方文字的距离

\caption{95\% Hausdorff distance of independent test set in the MICCAI 2015 dataset. Our approach has better 95HD because the enlarged receptive field can alleviate outliers.}\label{MICCAI_HD}

\end{table}

\clearpage
\begin{figure}[!ht]
\centering
  \subfigure{
    \includegraphics[width=0.22\linewidth]{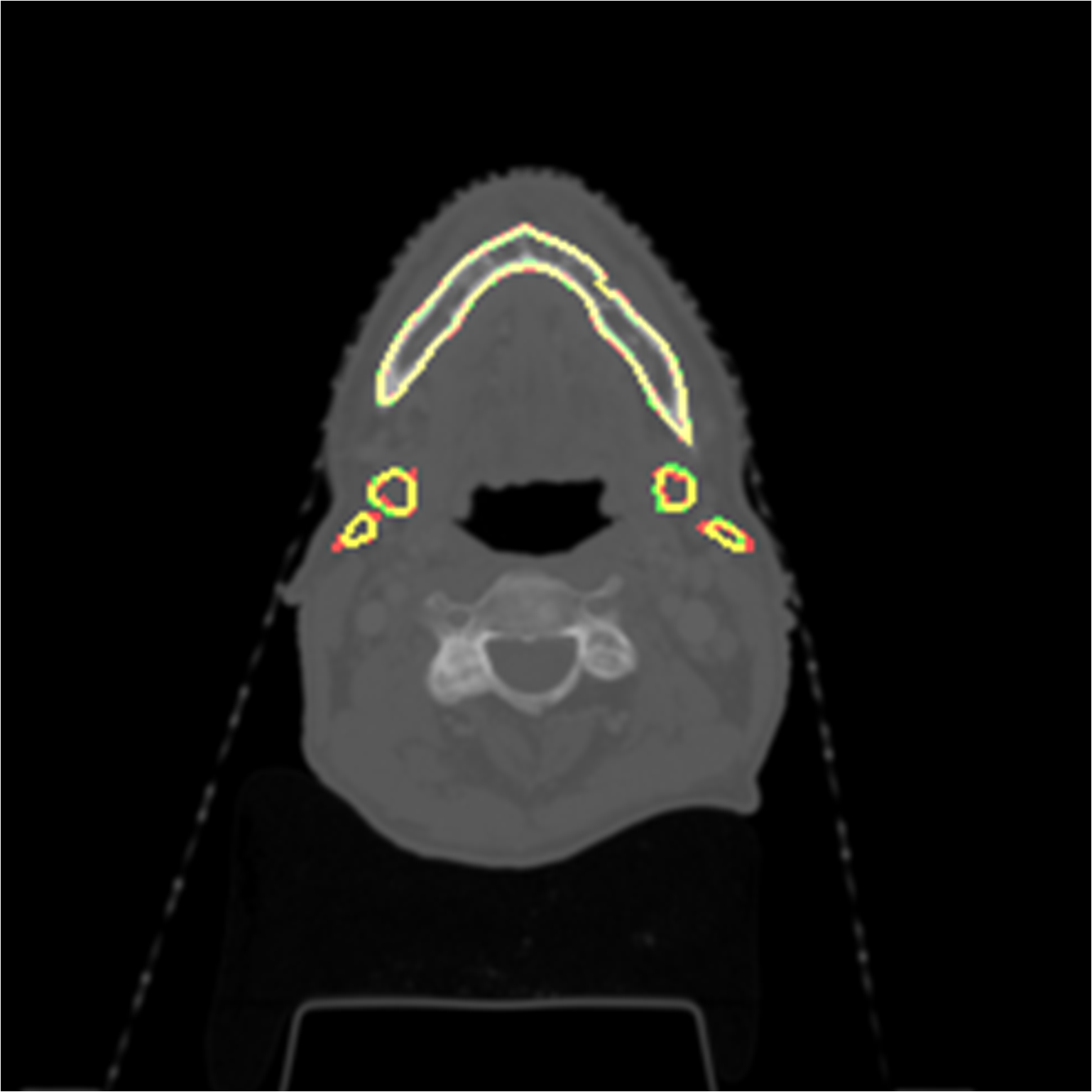}
  }
  \subfigure{
    \includegraphics[width=0.22\linewidth]{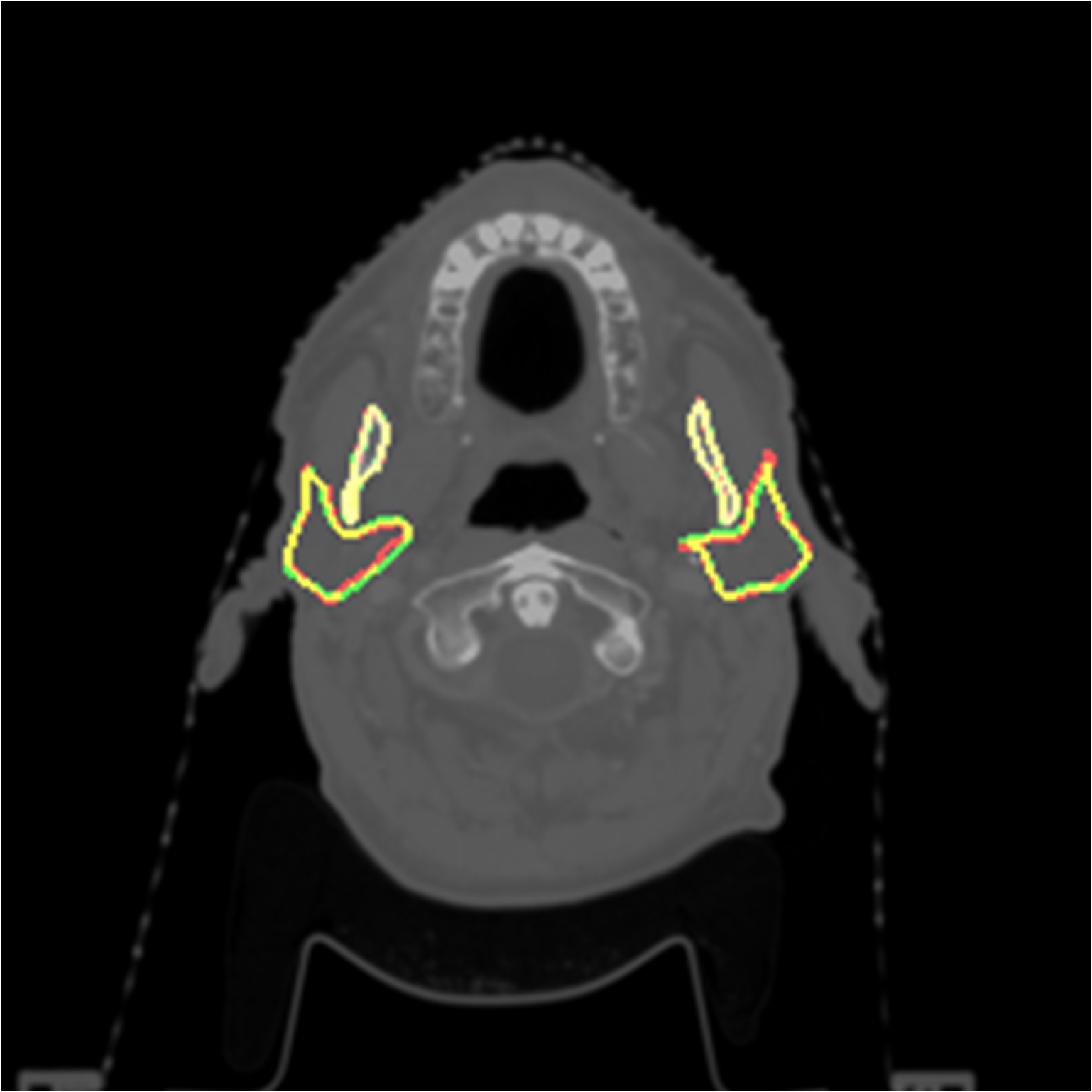}
  }
  \subfigure{
    \includegraphics[width=0.22\linewidth]{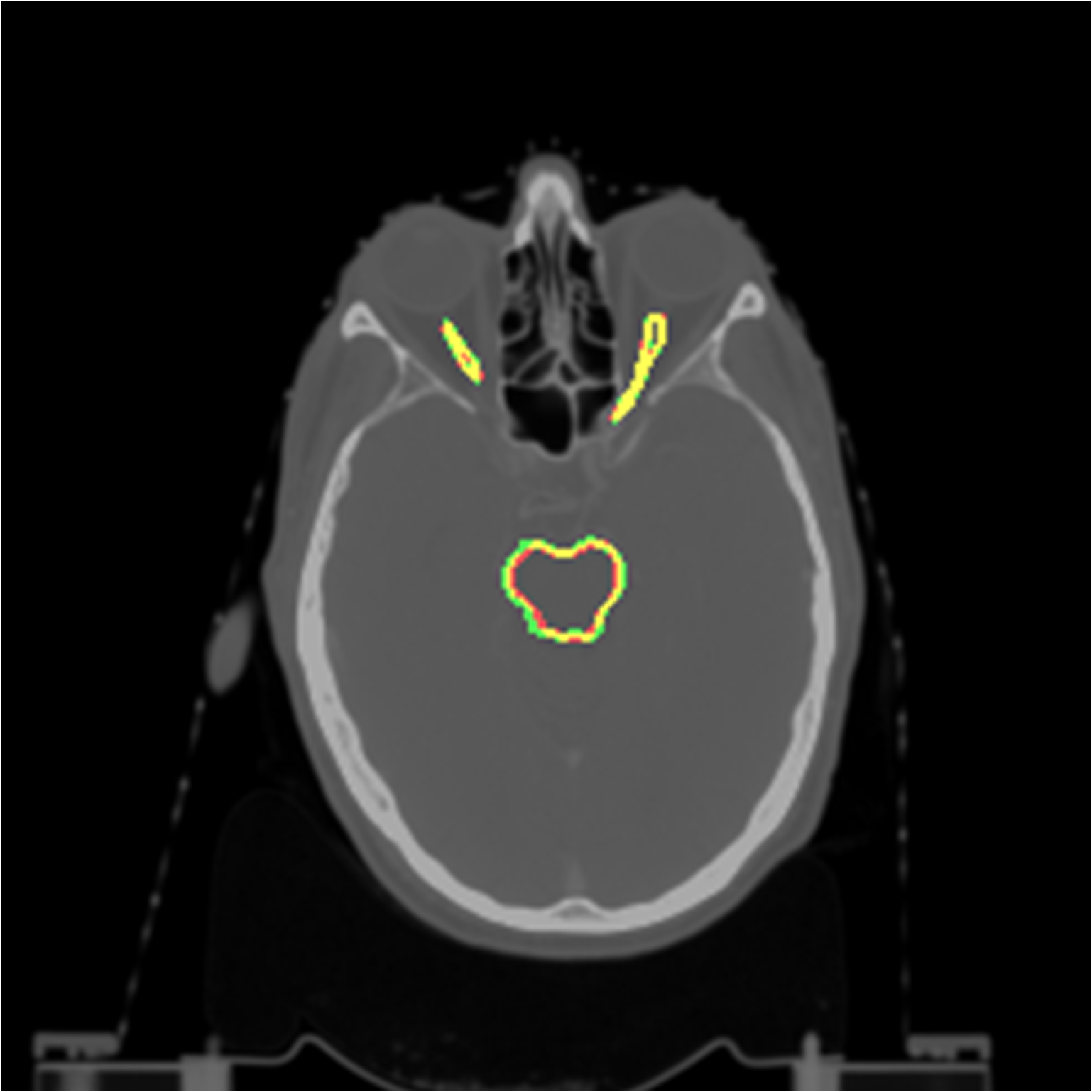}
  }
  \subfigure{
    \includegraphics[width=0.22\linewidth]{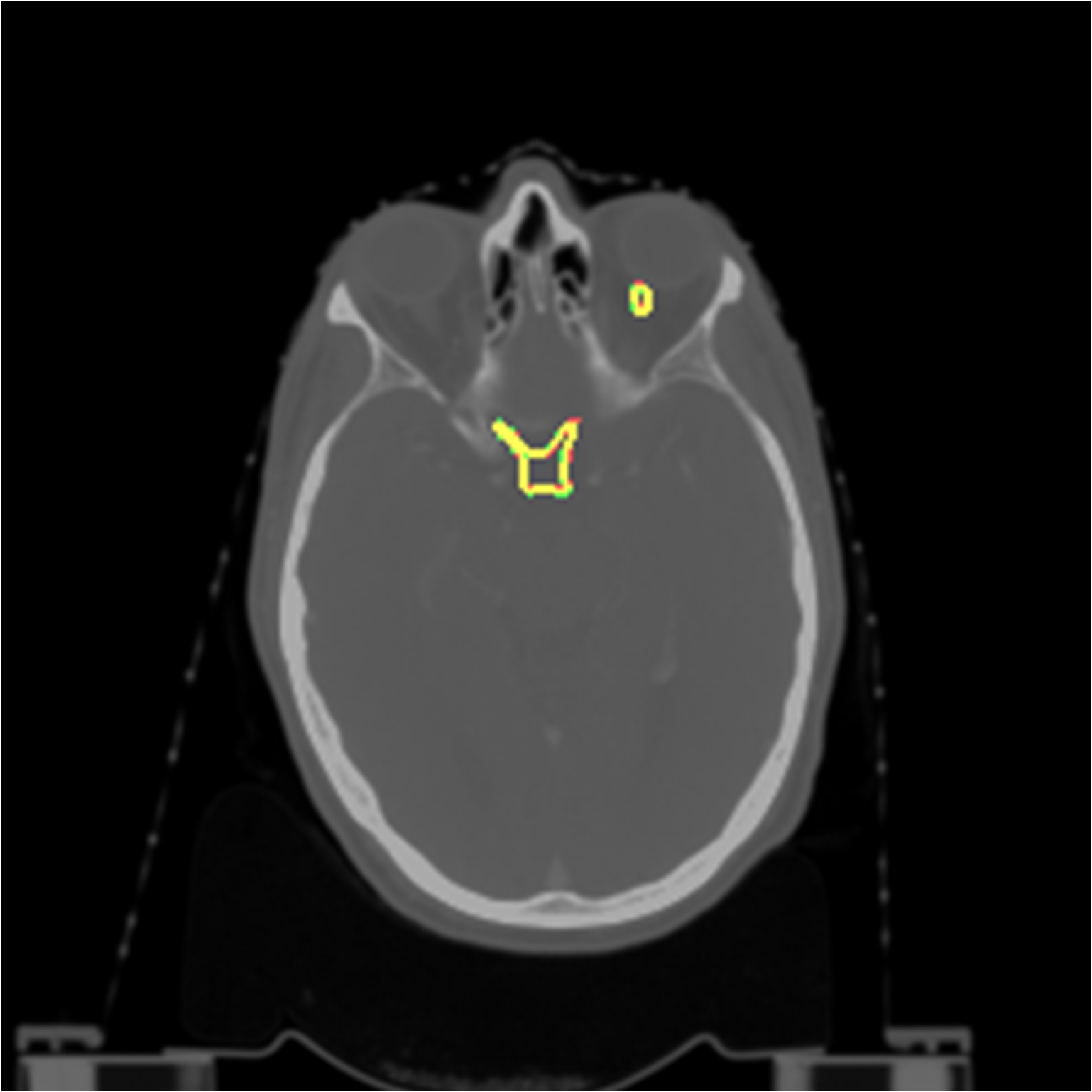}
  }\\
  \subfigure{
    \includegraphics[width=0.22\linewidth]{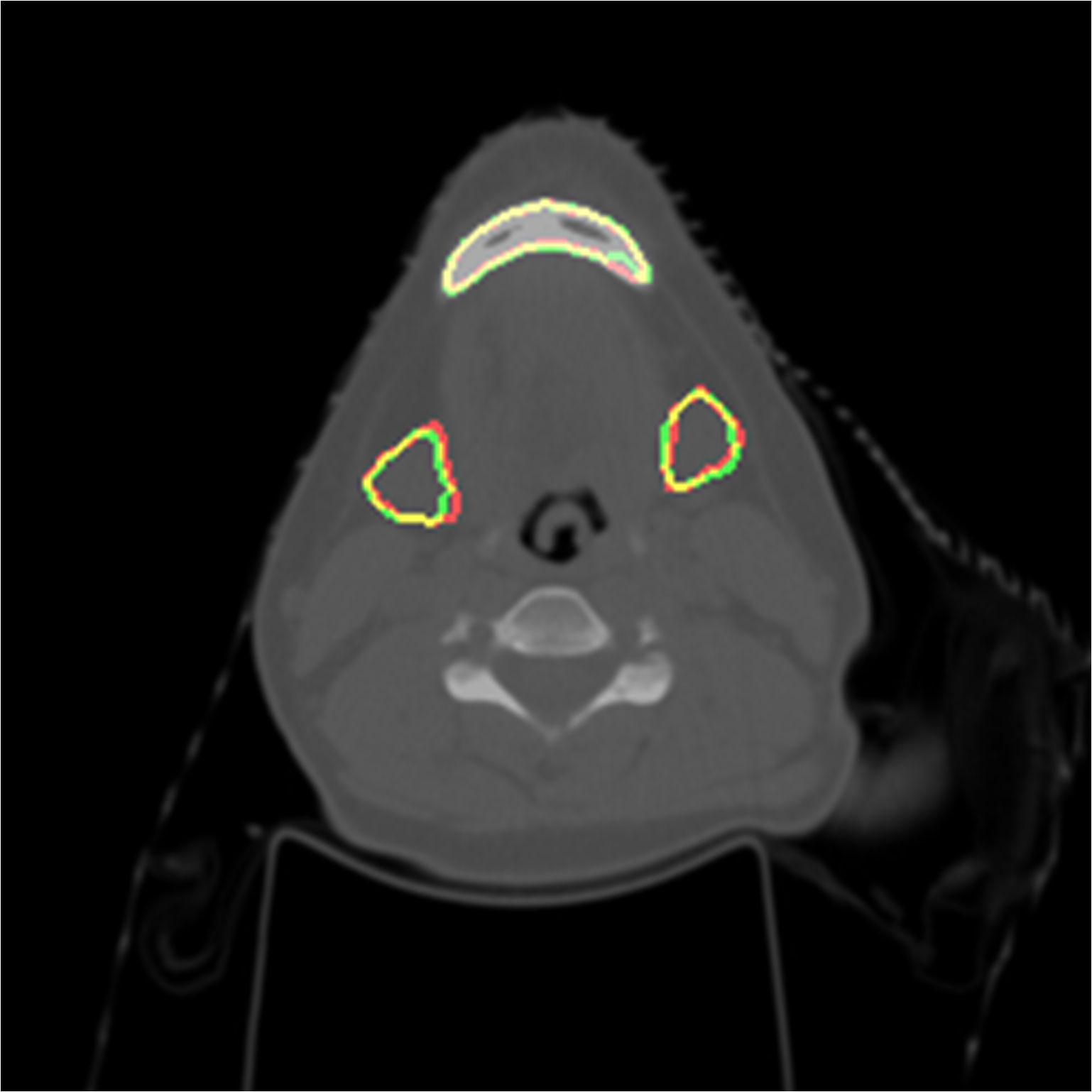}
  }
  \subfigure{
    \includegraphics[width=0.22\linewidth]{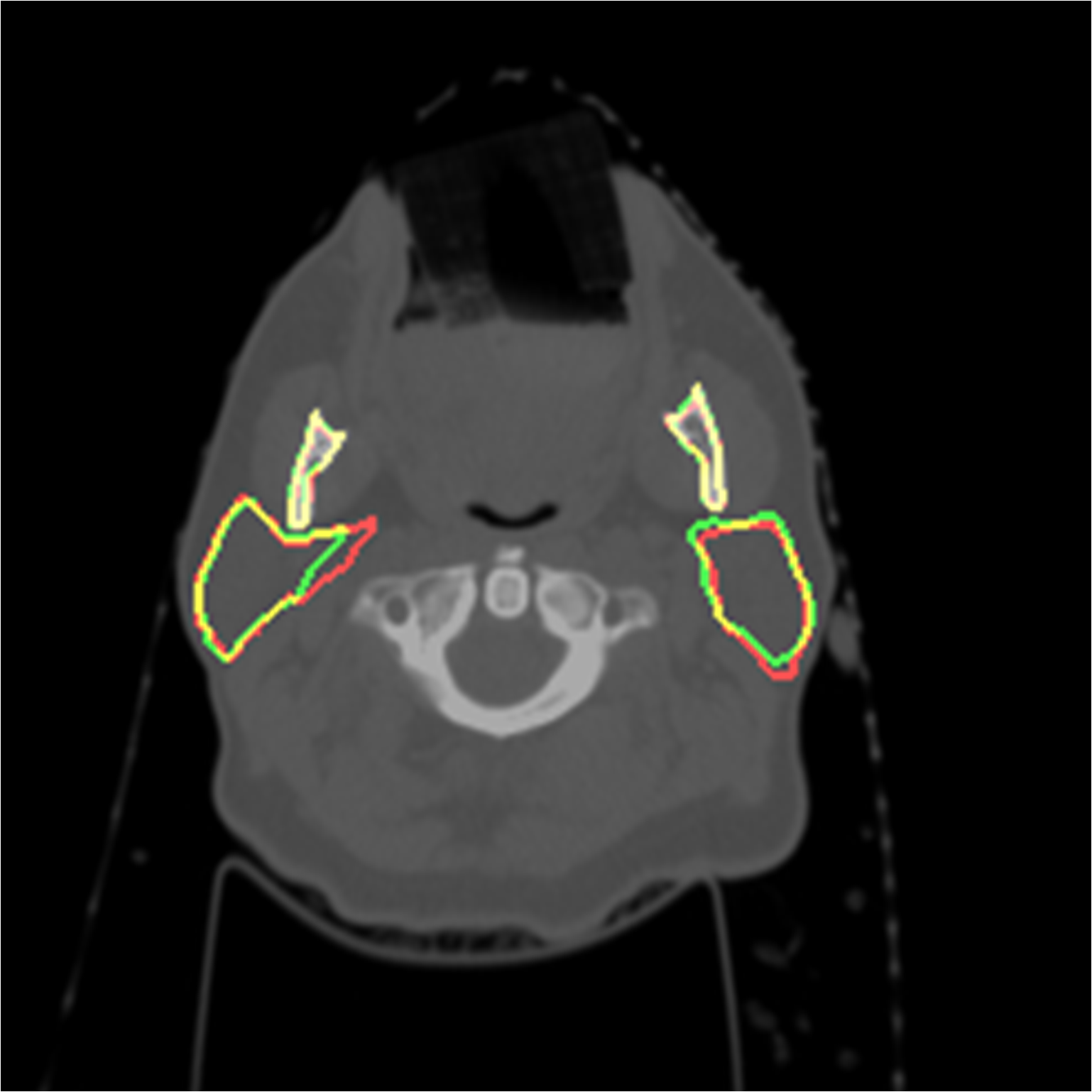}
  }
  \subfigure{
    \includegraphics[width=0.22\linewidth]{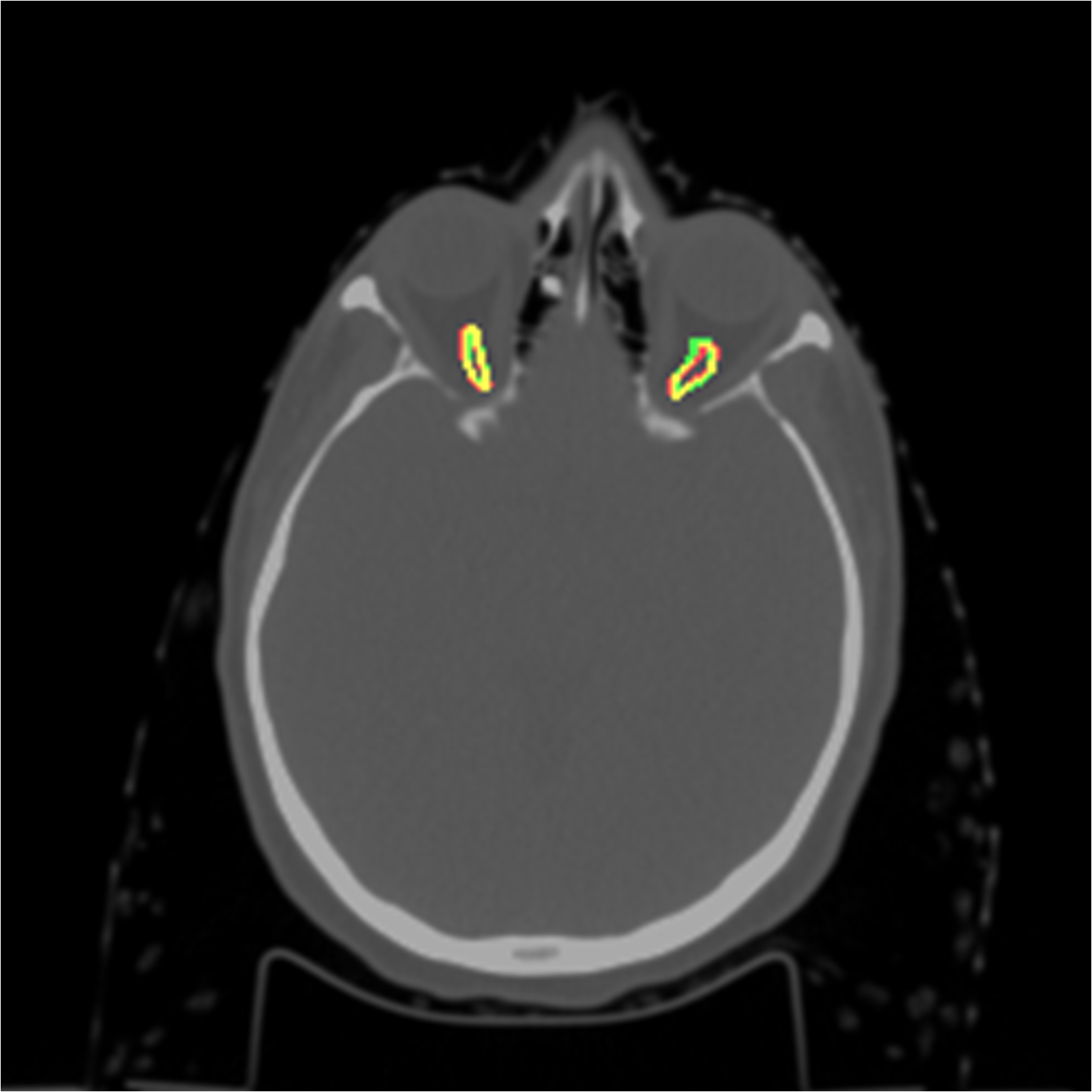}
  }
  \subfigure{
    \includegraphics[width=0.22\linewidth]{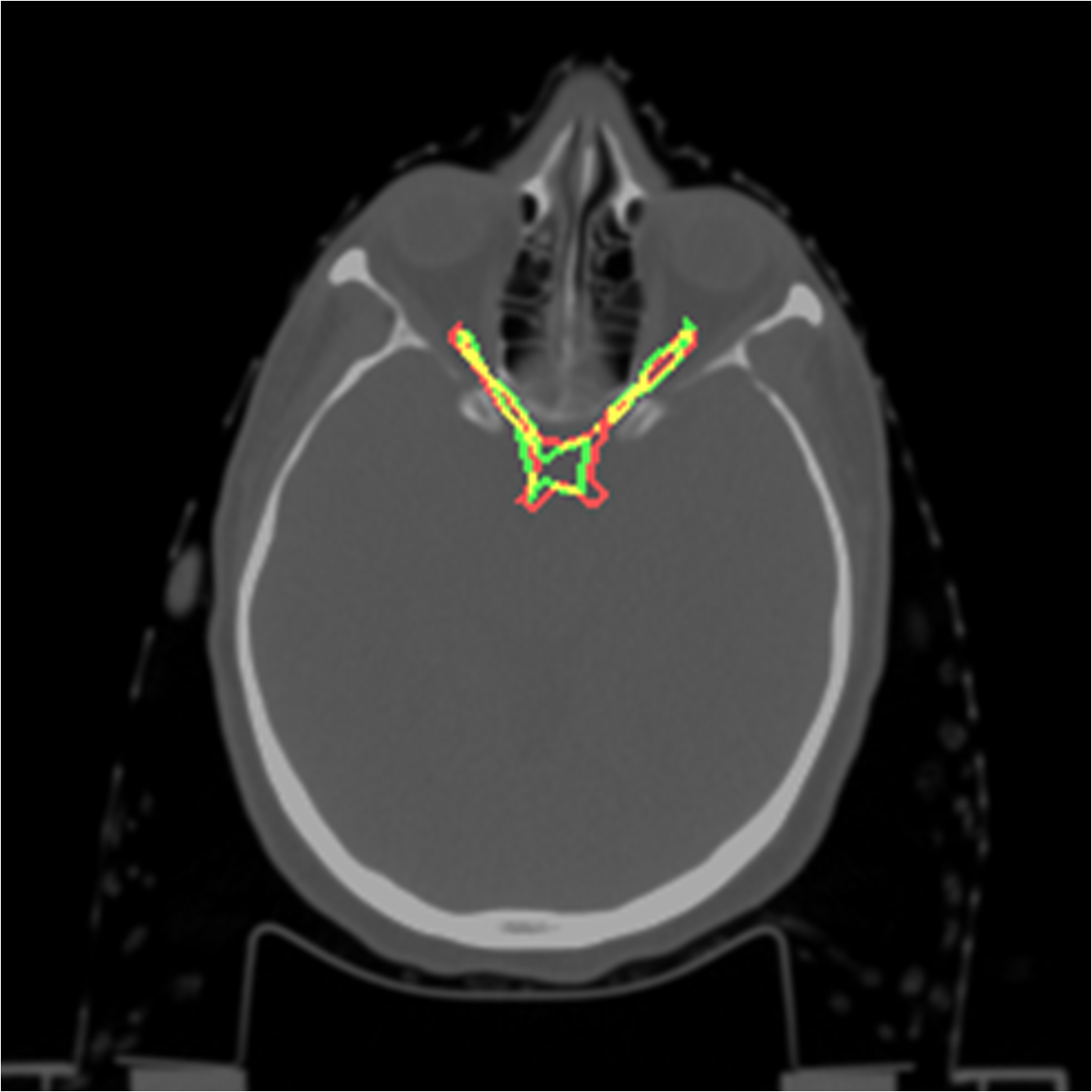}
  }\\
  \captionsetup{font={normal}}
  \caption{Segmentation results on sample 0522c0576 (first row), and 0522c0661 (second row) by our FocusNet. The green lines are the contour of predicted organ boundaries, red lines are the contour of ground truth, and the yellow lines are the overlap.}
  \label{fig:MICCAI_fig}
\end{figure}

\end{document}